\documentclass[conference]{IEEEtran}

\usepackage{graphicx} 
\usepackage{spverbatim}
\usepackage{enumitem}
\usepackage{booktabs, tabularx} 
\usepackage[caption=false]{subfig}
\usepackage{svg}       
\usepackage{placeins}
\usepackage{multirow}
\usepackage{array}
\usepackage{tikz}
\usepackage{tcolorbox}
\usepackage{xcolor} 
\usepackage{enumitem}
\usepackage{adjustbox}
\usepackage{wrapfig}
\usepackage{helvet}
\usepackage{wasysym}
\usepackage{float}
\usepackage{multicol}
\usepackage{url}

\definecolor{palebluegray}{rgb}{0.86, 0.92, 0.98}
\definecolor{palegreengray}{rgb}{0.82, 0.91, 0.78}
\definecolor{palerosegray}{rgb}{0.88, 0.79, 0.83}
\definecolor{palemelon}{rgb}{0.96, 0.68, 0.}
\definecolor{brightblue}{rgb}{0.85, 0.93, 0.99}

\newcommand{\textbox}[1]{%
\begin{tcolorbox}[colback=white, colframe=black, boxrule=0.5pt, left=.5mm, right=.5mm, top=.2mm, bottom=.5mm]
    \small #1
    \vspace{-3pt}
\end{tcolorbox}
}

\definecolor{dkgreen}{HTML}{006400}

\definecolor{yellow}{HTML}{FFFF00}

\newcolumntype{L}[1]{>{\raggedright\arraybackslash}p{#1}} 
\newcolumntype{C}[1]{>{\centering\arraybackslash}p{#1}} 
\newcolumntype{M}[1]{>{\centering\arraybackslash}m{#1}}

\newcommand{\boxedtext}[2][]{%
    \fcolorbox{gray!50}{white!95!gray}{%
        \parbox{0.9\linewidth}{%
            \centering
            \color{black}
            \ifx&#1&%
            \else
            \textbf{#1}\\[0.5em]
            \fi
             {\fontfamily{rm}\selectfont \textit{#2}} 
        }%
    }%
}

\title{ROOT: Requirements Organization and Optimization Tool}
\author{Katherine R. Dearstyne, Alberto D. Rodriguez and Jane Cleland-Huang\\University of Notre Dame\\ \{kdearsty, arodri39, JaneClelandHuang\}@nd.edu}
\date{October 2023}
\author{
\IEEEauthorblockN{Katherine R. Dearstyne}
\IEEEauthorblockA{\textit{Computer Science and Engineering} \\
\textit{University of Notre Dame}\\
Notre Dame, IN, USA \\
kdearsty@nd.edu}
\and
\IEEEauthorblockN{Alberto D. Rodriguez}
\IEEEauthorblockA{\textit{Computer Science and Engineering} \\
\textit{University of Notre Dame}\\
Notre Dame, IN, USA\\
arodri39@nd.edu}
\and
\IEEEauthorblockN{Jane Cleland-Huang}
\IEEEauthorblockA{\textit{Computer Science and Engineering} \\
\textit{University of Notre Dame}\\
Notre Dame, IN, USA \\
JaneClelandHuang@nd.edu}
}
\begin{document}
 \maketitle
\begin{abstract}
Software engineering practices such as constructing requirements and
establishing traceability help ensure systems are safe, reliable, and maintainable. However, they can be resource-intensive and are frequently underutilized. To alleviate the burden of these essential processes, we developed the Requirements Organization and Optimization Tool (ROOT). ROOT centralizes project information and offers project visualizations and AI-based tools designed to streamline engineering processes. With ROOT's assistance, engineers benefit from improved oversight and early error detection, leading to the successful development of software systems. \\ Link to screen cast: \url{https://youtu.be/3rtMYRnsu24}

\end{abstract}

\section{Introduction}
\label{sec:intro}
Collaboration is critical to the success of building any complex software system, yet the coordination of individuals from different backgrounds and expertise presents challenges of its own \cite{bertram_communication_nodate, yasrab_challenges_2019}. Practices such as constructing requirements, establishing traceability, and performing validation and testing are designed to mitigate potential issues earlier and facilitate better communication across individuals and groups \cite{whitehead_collaboration_2007, lehman_software_1991}. Despite the usefulness of these practices, they are often difficult to build into the process due to the additional time and resources required to properly implement them \cite{bayer_view-based_2006, ahonen_case_2003, lim_balancing_2012}. As a result, they are often ignored, delayed, or inadequately sustained \cite{DBLP:conf/icse/RempelMKC14, ghanbarihadi_omission_2018}, especially in startups and small companies where speed is often prioritized over comprehensive requirements engineering processes \cite{paternoster_software_2014, giardino_software_2016}. Furthermore, even when these practices are implemented, collaboration between groups can introduce inconsistencies or confusion that ultimately results in failures down the road \cite{curtis_field_1988}. Therefore, easing the burden of these processes on engineers is paramount to ensure systems are safe, reliable, and maintainable. 

\begin{figure}
    \centering
    \includegraphics[width=\columnwidth]{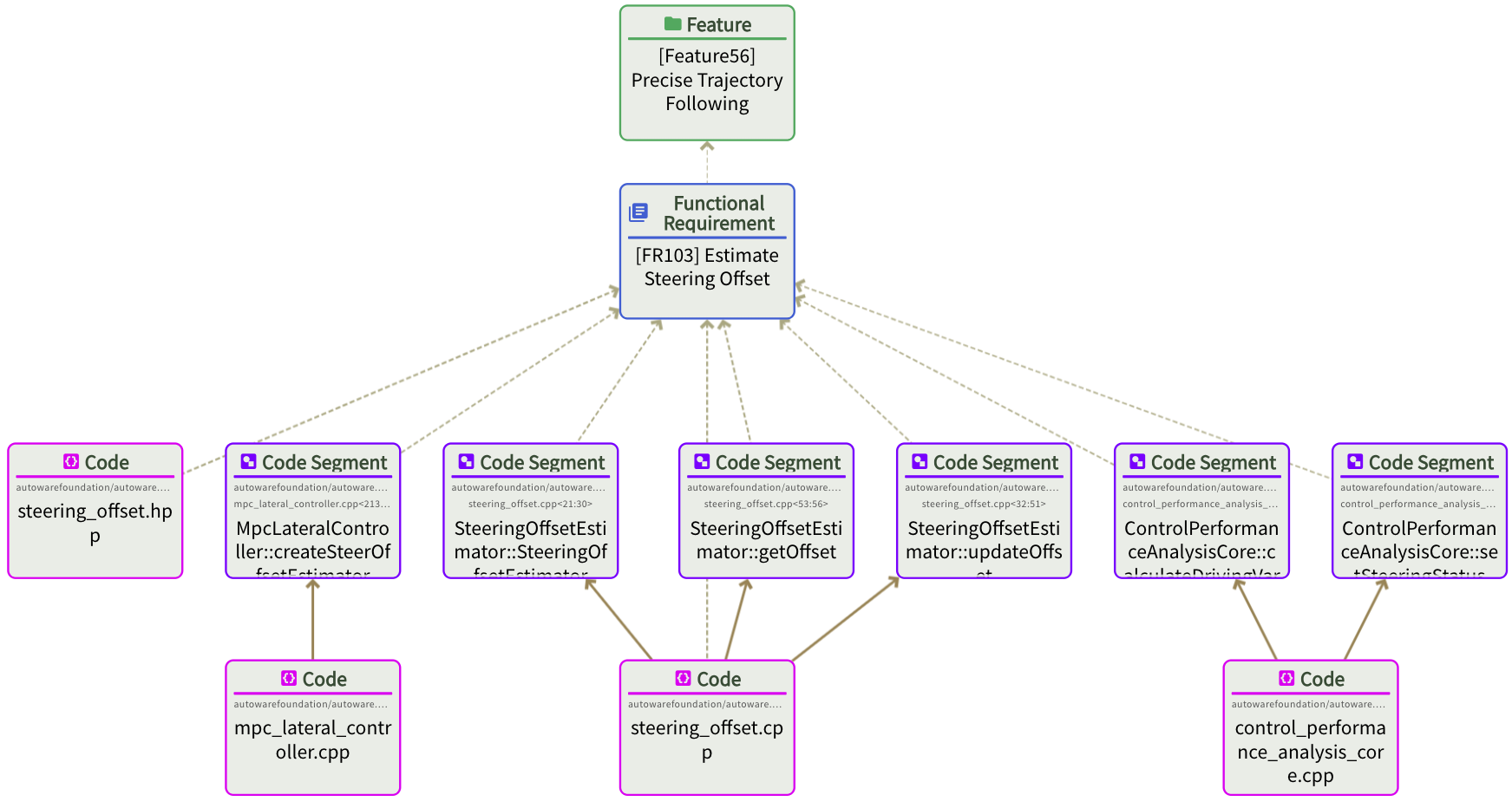}
    \caption{Slice of documentation and links generated for Autoware \cite{Autowarefoundation} open source project.}
    \label{fig:cover}
\end{figure}

To ease the burden of requirements and software engineering processes, we developed \textbf{ROOT} (\textbf{R}equirements \textbf{O}rganization and \textbf{O}ptimization \textbf{T}ool). ROOT serves as the hub of all project information, organized into a hierarchical artifact tree (Figure \ref{fig:cover}) that allows users to easily navigate their projects. Additionally, ROOT offers a range of tools designed to expedite engineering processes and provide engineers with an additional layer of oversight to catch mistakes early, acting as an \emph{assistant} rather than a replacement for human engineers. These tools encompass the following core features:

\begin{itemize}
\item \textbf{Integration Across Knowledge Sources:} ROOT integrates seamlessly with GitHub, Jira, and other project knowledge sources, centralizing all project information within the platform. \label{feature:integration}
\item \textbf{Project Summarization:} ROOT automatically generates a summary of the software project, highlighting sub-systems and project features. It also offers summaries of individual source code files to accelerate project understanding. \label{feature:summarization}
\item \textbf{Artifact Creation/Generation:} Engineers can load existing artifacts into ROOT or create new ones through a user-friendly interface. When documentation is lacking, ROOT automatically generates hierarchical layers of documentation from source code.\label{feature:generation}
\item \textbf{Project Visualization} Users can visualize their project in multiple ways, including an artifact tree or a table. To focus on a specific aspect at at time, they can select a subset of their project and create a \emph{view} of it. \label{feature:visualization}
\item \textbf{AI-based Project Chat:} ROOT provides a chat interface for project-specific queries, which references any pertinent artifacts in its responses, enabling users to verify information easily.\label{feature:chat}
\item \textbf{Trace-link Generation:} ROOT predicts trace links between related artifacts, providing explanations for each link. Reviewers can easily accept or reject predicted links and create new ones.\label{feature:tracelinks}
\item \textbf{Project Vocabulary:} To encourage consistent terminology across the project, ROOT automatically extracts project concepts and vocabulary and links them to the relevant artifacts. \label{feature:vocab}
\item \textbf{Requirements Assessment:} ROOT automatically checks requirements for inconsistencies or ambiguities. Reviewers can also manually flag requirements to alert the relevant parties.\label{feature:requirements}
\end{itemize}
The remainder of this paper is organized as follows: Section \ref{sec:related} presents previous work that ROOT builds upon. Section \ref{sec:architecture} outlines the technical details of the tool. Section \ref{sec:safa} performs a walk through of each features in more detail. Finally, Section \ref{sec:future_work} contains future work and Section \ref{sec:conclusion} concludes our paper.

\section{Related Work}
\label{sec:related}

Early management systems like DOORs focused on artifact management and trace link organization, laying the groundwork for representing engineering processes in software. Over time, tools like JIRA and Siemens's Polarion expanded their roles, with JIRA evolving from issue tracking to managing the entire software development process, and Polarion addressing system requirements. While these tools were manual and labor-intensive, modern alternatives like JAMA, Visure Solutions, Osseno, and SpiraTeam have begun integrating AI for enhancements such as quality checks, onboarding assistance, and the generation of user stories and test cases, though they primarily augment existing processes rather than creating new ones. ROOT distinguishes itself by catering to startups and small companies that often lack rigorous software management in the early stages of product development. Its documentation generation pipeline produces an initial draft of system requirements linked to source code, significantly reducing the effort of documentation creation and allowing teams to focus on refinement. Furthermore, through collaborations with NASA's Goddard and Jet Propulsion Laboratories, ROOT has identified additional features that can help multi-team organizations detect conflicts early and stay aligned throughout the project life-cycle. These include identifying inconsistencies across an entire project’s requirements and maintaining a centralized graph of project concepts.  Collectively, these capabilities enable ROOT to support organizations at all stages of maturity.

\section{Architecture and Processes}
\label{sec:architecture}
\subsection{Architecture}

The ROOT system consists of three key components: a front-end application (FEND), a back-end server (BEND), and a generation server (GEN). FEND, built with Vue.js and Cytoscape, visualizes projects, artifacts, and trace links as interactive graphs, with options to view these elements in tables or query them via a chat interface. BEND, implemented in Java with Spring Boot, manages user interactions, data retrieval, and job execution through a REST API, using a MySQL database to store all project-related data. GEN, a Python-based Django application with Celery for job management, handles AI-driven tasks like documentation generation and chat features, leveraging advanced language models and providing updates on job status and results. Together, these components create a scalable system for managing and generating project documentation and traceability.

\subsection{Job-Execution}





In ROOT, job execution begins when a user initiates a job through the front-end client, prompting the back-end server to create a job entry, start execution, and provide a job ID for tracking. The back end then asynchronously retrieves project data and sends it to the generation server (GEN), which processes the job and returns its ID for progress updates. Throughout the job, the back end polls GEN for updates, logs progress, and, upon completion, retrieves and processes results from an S3 bucket. The back end then saves new artifacts and trace links, marks the job as complete, and notifies the user via email and WebSockets, ensuring efficient and automated requirements engineering tasks.

\subsection{AI-based Features}
\label{sec:ai-features}

ROOT's value lies in its comprehensive feature set, built upon existing research which we link in case more details are wanted. For tasks like summarization, document generation, chat, and explanations, ROOT leverages Anthropic's LLMs \cite{claude}, known for their large context windows, and enhances their output with techniques such as Retrieval-Augmented Generation (RAG \cite{lewis_retrieval-augmented_2021}), Chain-of-Thought Reasoning \cite{wei_chain--thought_2023}, and the ReAct approach \cite{yao_react_2023}. For example, ROOT's chat and inconsistency detection methodologies are similar to those described in \cite{ezzini_ai-based_2023}, while our trace-link explanations are based on methods used in \cite{rodriguez_prompts_2023}. Additionally, ROOT employs cross-encoder and embedding models, including BERT variants \cite{bert_pl, bert_nl} and Sentence-BERT \cite{reimers_sentence-bert_2019}, for tasks like trace-link prediction and clustering artifacts in the documentation generation pipeline.

\section{Walk Through}
\label{sec:safa}

This section provides a graphical walkthrough of ROOT, illustrating the user experience and interactions with the system for each of the core features outlined in Section \ref{sec:intro}. Throughout our walkthrough, we use a subset of the Autoware Foundation open-source project \cite{githubGitHubAutowarefoundationautowareuniverse} as a running example.

\begin{figure}
    \centering
    \includegraphics[width=\columnwidth]{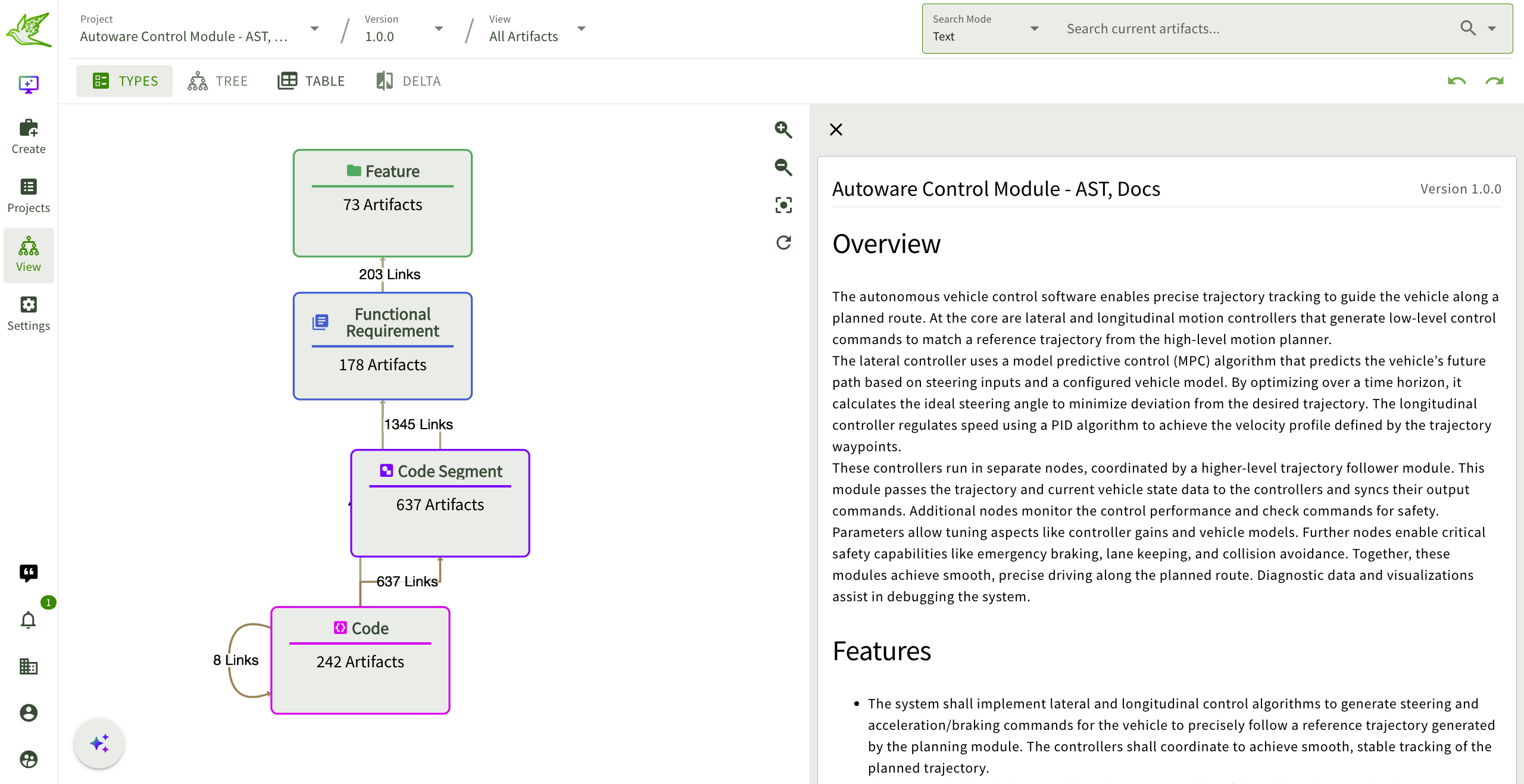}
    \caption{Project overview screen containing Traceability Information Model (TIM) and graph controls. Project summary excluded for display purposes.}
    \label{fig:project_overview}
\end{figure}

\subsection{On-boarding:}
Upon entering the tool, users are encouraged to go through an on-boarding process to help them import their project in the platform. We have observed that many users possess only code and want to take advantage of ROOT’s features designed to address gaps in their software engineering practices. Therefore, this initial on-boarding is specifically tailored for users whose primary data source is GitHub. Users with different needs can opt out of this workflow and start from the project import screen which allows for more flexibility.

During on-boarding, users are prompted to import their project codebase, generate a project summary, and create any missing documentation. 

\subsection{Integration Across Knowledge Sources} From the project import screen, users are given three ways to import a project: via GitHub, Jira, or flat files. If a user is beginning a new project, they may choose to begin with a blank project.

If a user opts to import an existing project, they are prompted to connect their GitHub or Jira account, or to upload the required files. Once this is done, an import job is initiated, allowing the user to monitor its progress. 

For the remainder of the walk-through, we will assume that the user is connecting their Git-hub account and move through the remaining on-boarding steps.

 \subsection{Project Summarization} 
 Once a user has imported their project, a summary of the project will be automatically generated, along with summaries for all code files. The project summary includes an overview of the system and sections detailing sub-systems, entities, features, and data flow.


\begin{figure}[b]
    \centering
    \includegraphics[width=\columnwidth]{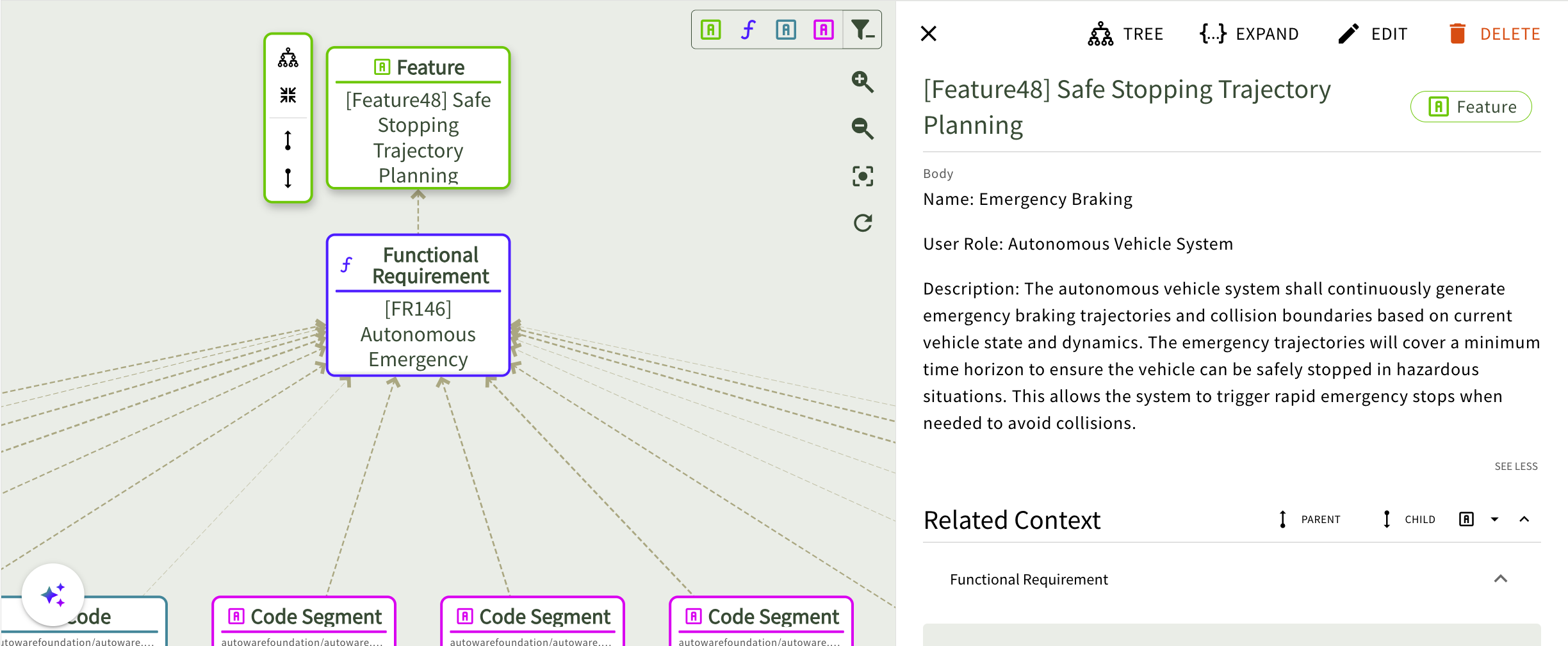}
    \caption{Selected artifact view of generated feature containing description and further details.}
    \label{fig:artifact_selection}

\end{figure}

\begin{figure}[t]
    \centering
    \includegraphics[width=.95\columnwidth]{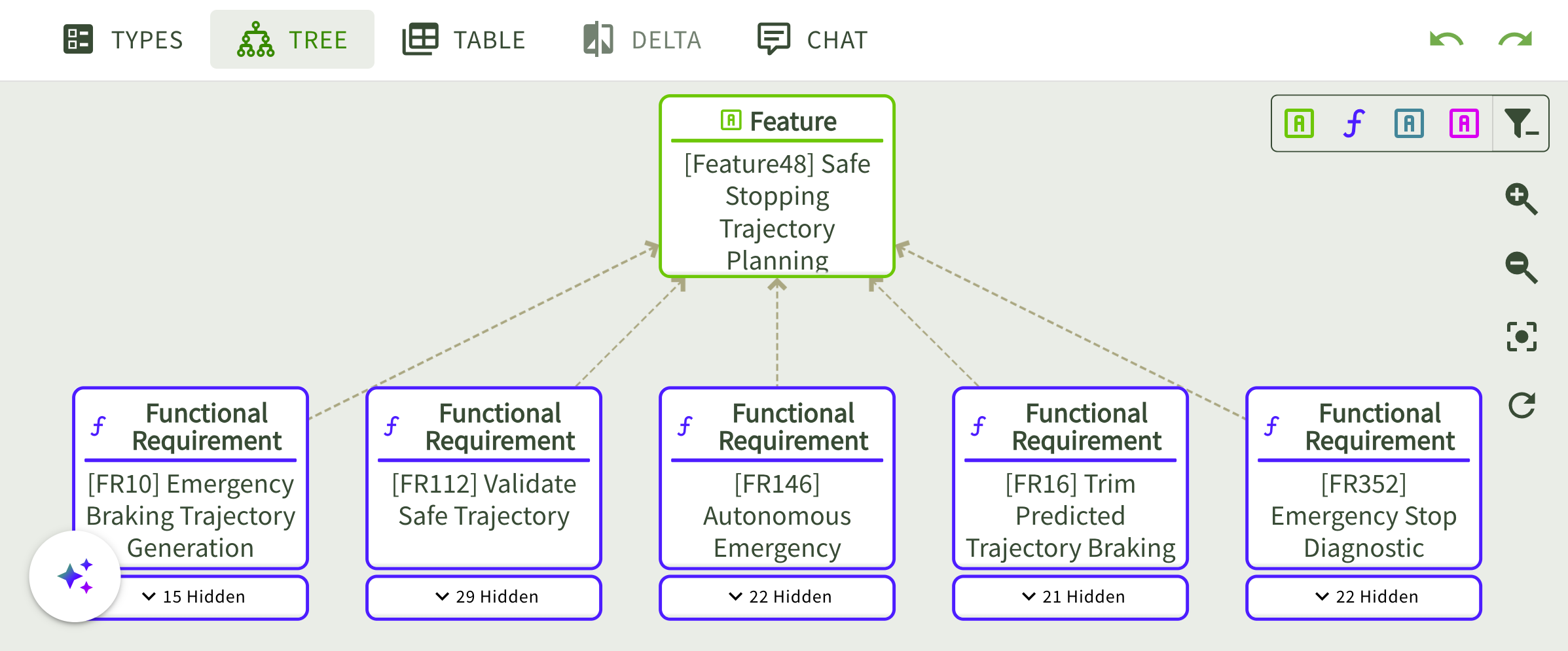}
    \caption{Tree view of generated feature from Autoware Project.}
    \label{fig:artifact_tree}
\end{figure}

\begin{figure}[t]
    \centering
    \includegraphics[width=.95\columnwidth]{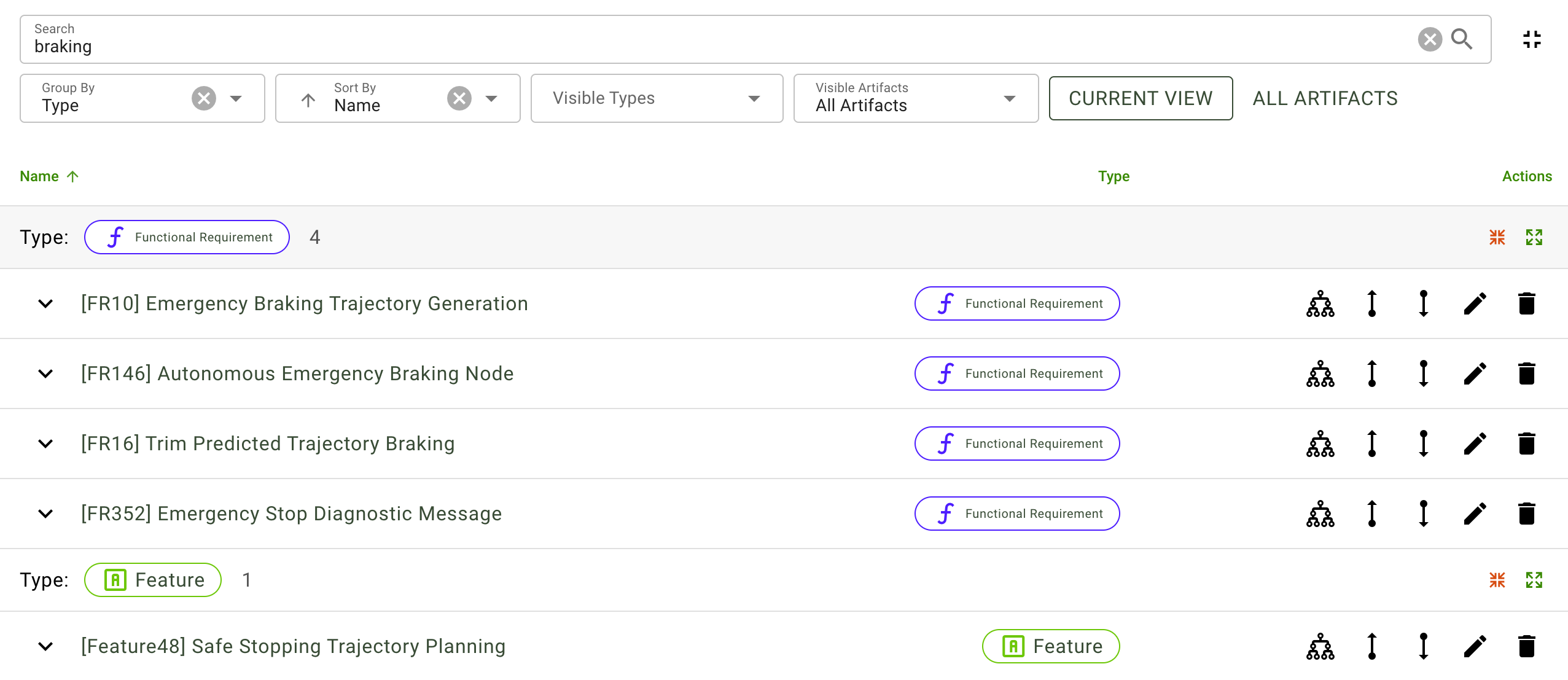}
    \caption{Table view of search results for query 'braking'. }
    \label{fig:artifact_table}
\end{figure}

\subsection{Artifact Generation}
The on-boarding process also allows users to generate two new layers of documentation for their project. If they choose to do so, the system initiates a multi-step pipeline that was developed and evaluated in previous work \cite{dearstyne_supporting_2024}. This pipeline first identifies groups of related code based on common functionality, then uses a LLM to generate Functional Requirements and then repeats the process to generate higher-level Features for them. While ROOT can also generate various document formats, such as user stories or design documents, we found that Functional Requirements and Features are particularly beneficial for users when they are first starting out. However, once the users enter the platform, they can generate additional types of documentation at any time.

\subsection{Project Visualization}
After onboarding, users are directed to the project home page (as shown in Figure \ref{fig:project_overview}), which displays the project's Traceability Information Model (TIM), a generated project summary, and graph controls. This overview helps users quickly grasp the project's structure and key features. ROOT offers two primary methods for viewing artifacts: a tree view (Figure \ref{fig:artifact_tree}) and a table view (Figure \ref{fig:artifact_table}) . For large projects, the tree visualization is disabled by default to avoid overwhelming users, though they can target a single artifact to view in the graph. The tree view also allows for expanding and collapsing branches to manage complexity. The table view, suitable for handling numerous artifacts, enables filtering, sorting, and fuzzy searching, with an additional tab for viewing and managing trace links, including their review status.

When users click on an artifact in either view, a side panel opens with more detailed information (Figure \ref{fig:artifact_selection}), and double-clicking allows them to create a focused \emph{view} centered on the selected artifact and its relationships. This setup is demonstrated using the Autoware project example, where users can efficiently locate and explore specific features, such as the braking functionality, and visualize its parent and children. A video walkthrough provides further details on these visualization capabilities, offering users a practical guide to navigating their projects within ROOT.

\begin{figure}
    \centering
    \includegraphics[width=.75\columnwidth]{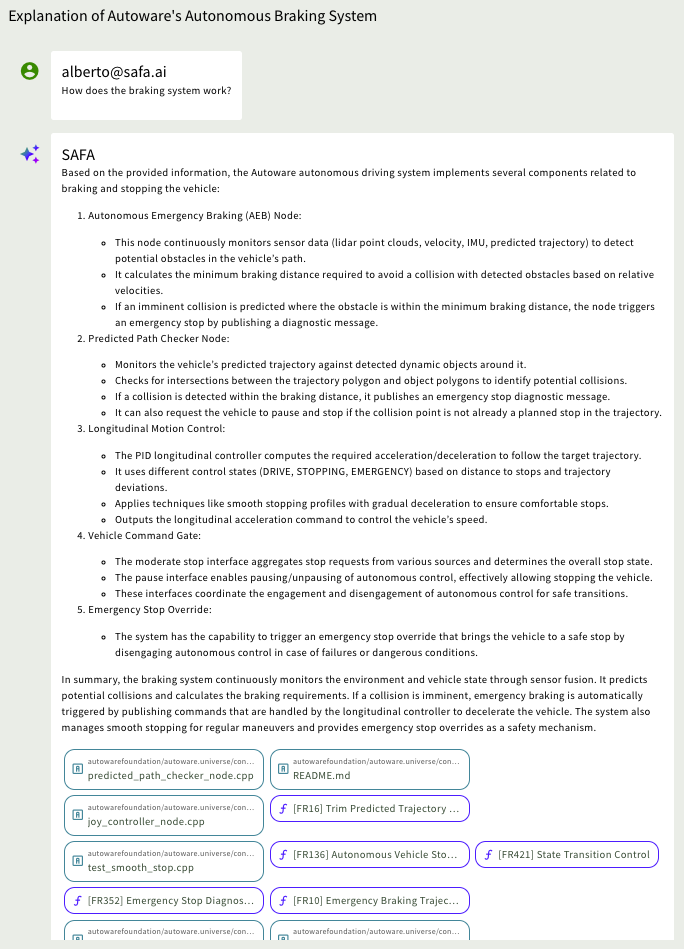}
    \caption{Example chat response for Autoware project braking sub-system.}
    \label{fig:chat}
\end{figure}


%


\subsection{AI-based Project Chat:} 
Alongside the tree and table view buttons, users can access the chat interface. From this screen, users can ask questions about their project and receive responses generated by a LLM, as shown in Figure \ref{fig:chat}. To encourage accurate responses, the LLM is provided with related project artifacts. These artifacts are displayed to the user as clickable buttons beneath the LLM's response. Clicking on one of these artifacts opens a detailed view of that artifact.
\vspace{-1.2mm}
\subsection{Trace-link Generation:} 
\begin{figure}[b]
    \centering
    \includegraphics[width=.95\columnwidth]{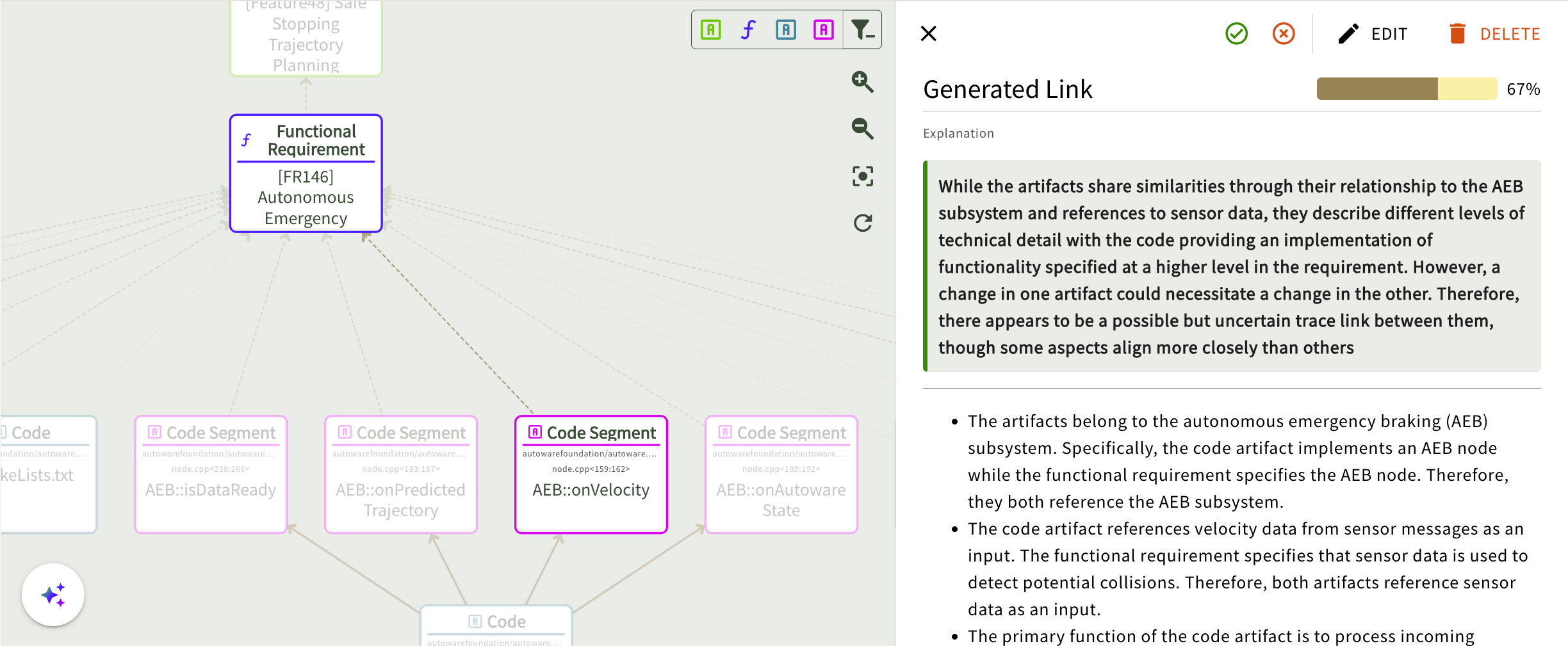}
    \caption{Detailed view of generated link between generated documentation and a code segment.}
    \label{fig:trace_selection}
\end{figure}

To capture any potentially missed relationships, ROOT provides automatic trace-link generation. Users can select the child type(s) and parent type(s) for which trace-links will be predicted. Once the predictions are completed, all generated links will be added to the project as dotted lines for user review. Additionally, a confidence score and an LLM-generated explanation will be provided to elaborate on the potential relationship between the artifacts as can be seen in Figure \ref{fig:trace_selection}.

\subsection{Project Vocabulary:} 
ROOT also assists users in maintaining a project vocabulary to facilitate better alignment among different experts involved in a project. Terminology is automatically identified during health checks, as detailed in the following section. Additionally, users can manually add new concepts at any time, similar to adding a traditional artifact. These concepts are then utilized as context when viewing artifacts, querying in the chat, or evaluating other requirements.

\subsection{Requirements Assessment:} 
Users can perform a \emph{health check} on any of their natural language documents to identify potential issues with requirements.  During the \emph{health check}, the system flags inconsistencies with other requirements and highlights undefined or ambiguous concepts within the current project vocabulary. If any terminology in the artifact is undefined, users can address the warning by adding the term to the project vocabulary. Furthermore, the system identifies any existing project terminology that is referenced in the artifact, with automatic links created between those concepts and the current artifact. 

To illustrate these health checks, we have constructed a small project containing 4 requirements and 2 project concepts (\emph{Job} and \emph{Database Entity}) \cite{rodriguez_2024_12574870}. For illustration, two of the requirements are shown below:

\textbox{\textbf{R1:} The system shall be able to save some entities to the database, perform a job, and return the result of the job to the user in under 1 minute.}
\textbox{\textbf{R4:} Saving entities to the database shall take between 0.5 seconds to 5 seconds to complete. }

\begin{figure}
    \centering
\includegraphics[width=.75\columnwidth]{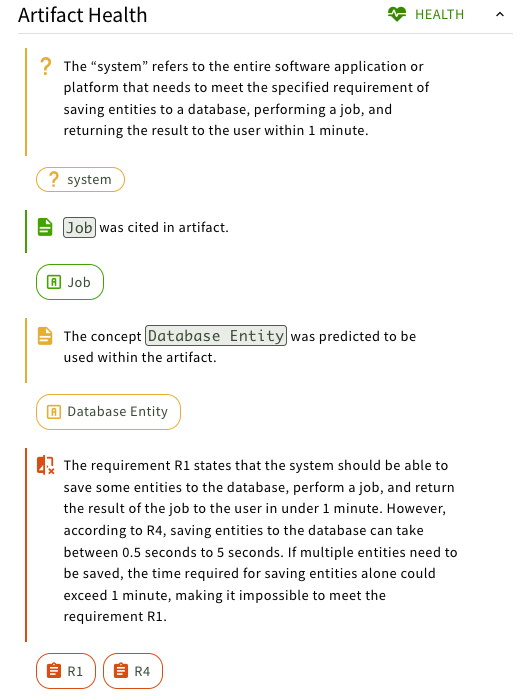}
    \caption{Artifact health checks showing cited concept, undefined concept, predicted concept, and contradiction in artifact `R1'.}
    \label{fig:concept_checks}
\end{figure}

If we run a health check on Requirement \emph{R1}, ROOT identifies several issues and display them  for resolution (Figure \ref{fig:concept_checks}).  The term \emph{system} has been flagged as a missing concept in the project vocabulary. An existing concept, \emph{Job}, is directly cited in the artifact, so the system creates a link between them. Additionally, the concept \emph{Database Entity}, although not directly cited, is predicted to be used in \emph{R1}. Finally, a contradiction is detected between \emph{R1} and \emph{R4} (see below) and an explanation is provided.

%

\section{Future Work}
\label{sec:future_work}
Our ultimate vision for ROOT is to support all aspects of the software engineering process comprehensively. To achieve this, we anticipate several future enhancements to ROOT. Firstly, we plan to integrate additional data sources, such as DOORs and PDFs, allowing engineers to connect all elements of their projects seamlessly. Additionally, we aim to enable users to link back to other tools, such as opening a code artifact in their preferred IDE. We also seek to improve artifact maintenance by identifying outdated documentation and updating it automatically. Finally, we intend to implement more health checks, such as flagging incomplete or unverifiable requirements. With these enhancements, ROOT will be better positioned to streamline software workflows and enhance overall project quality.

\section{Conclusion}
\label{sec:conclusion}
In this paper, we introduce ROOT, a tool designed to enhance collaboration and requirements management in software development. By automatically generating documentation and trace-links from the ground up, ROOT makes robust requirements engineering processes accessible even to startups and small companies. Through its visualizations and AI-based features, ROOT promotes more effective collaboration and early problem detection. Ultimately, ROOT aims to ease the burden of engineering processes and aid in developing safer, more reliable systems faster.

\bibliographystyle{IEEEtran}
\bibliography{software.bib}

\begin{thebibliography}{10}
\providecommand{\url}[1]{#1}
\csname url@samestyle\endcsname
\providecommand{\newblock}{\relax}
\providecommand{\bibinfo}[2]{#2}
\providecommand{\BIBentrySTDinterwordspacing}{\spaceskip=0pt\relax}
\providecommand{\BIBentryALTinterwordstretchfactor}{4}
\providecommand{\BIBentryALTinterwordspacing}{\spaceskip=\fontdimen2\font plus
\BIBentryALTinterwordstretchfactor\fontdimen3\font minus \fontdimen4\font\relax}
\providecommand{\BIBforeignlanguage}[2]{{%
\expandafter\ifx\csname l@#1\endcsname\relax
\typeout{** WARNING: IEEEtran.bst: No hyphenation pattern has been}%
\typeout{** loaded for the language `#1'. Using the pattern for}%
\typeout{** the default language instead.}%
\else
\language=\csname l@#1\endcsname
\fi
#2}}
\providecommand{\BIBdecl}{\relax}
\BIBdecl

\bibitem{bertram_communication_nodate}
D.~Bertram, A.~Voida, S.~Greenberg, and R.~Walker, ``\BIBforeignlanguage{en}{Communication, {Collaboration}, and {Bugs}: {The} {Social} {Nature} of {Issue} {Tracking} in {Small}, {Collocated} {Teams}}.''

\bibitem{yasrab_challenges_2019}
\BIBentryALTinterwordspacing
R.~Yasrab, J.~Ferzund, and S.~Razzaq, ``Challenges and issues in collaborative software developments,'' Mar. 2019, arXiv:1904.00721 [cs]. [Online]. Available: \url{http://arxiv.org/abs/1904.00721}
\BIBentrySTDinterwordspacing

\bibitem{whitehead_collaboration_2007}
\BIBentryALTinterwordspacing
J.~Whitehead, ``Collaboration in {Software} {Engineering}: {A} {Roadmap},'' \emph{Future of Software Engineering (FOSE '07)}, pp. 214--225, May 2007, conference Name: Future of Software Engineering ISBN: 9780769528298 Place: Minneapolis, MN, USA Publisher: IEEE. [Online]. Available: \url{http://ieeexplore.ieee.org/document/4221622/}
\BIBentrySTDinterwordspacing

\bibitem{lehman_software_1991}
\BIBentryALTinterwordspacing
M.~M. Lehman, ``\BIBforeignlanguage{en}{Software engineering, the software process and their support},'' \emph{\BIBforeignlanguage{en}{Software Engineering Journal}}, vol.~6, no.~5, pp. 243--258, Sep. 1991, publisher: IET Digital Library. [Online]. Available: \url{https://digital-library.theiet.org/content/journals/10.1049/sej.1991.0028}
\BIBentrySTDinterwordspacing

\bibitem{bayer_view-based_2006}
J.~Bayer and D.~Muthig, ``A view-based approach for improving software documentation practices,'' Apr. 2006, p. 10 pp.

\bibitem{ahonen_case_2003}
J.~Ahonen and T.~Junttila, ``A case study on quality-affecting problems in software engineering projects,'' Dec. 2003, pp. 145--153.

\bibitem{lim_balancing_2012}
E.~Lim, N.~Taksande, and C.~Seaman, ``A {Balancing} {Act}: {What} {Software} {Practitioners} {Have} to {Say} about {Technical} {Debt},'' \emph{Software, IEEE}, vol.~29, pp. 22--27, Nov. 2012.

\bibitem{DBLP:conf/icse/RempelMKC14}
\BIBentryALTinterwordspacing
P.~Rempel, P.~M{\"{a}}der, T.~Kuschke, and J.~Cleland{-}Huang, ``Mind the gap: assessing the conformance of software traceability to relevant guidelines,'' in \emph{36th International Conference on Software Engineering, {ICSE} '14, Hyderabad, India - May 31 - June 07, 2014}, P.~Jalote, L.~C. Briand, and A.~van~der Hoek, Eds.\hskip 1em plus 0.5em minus 0.4em\relax {ACM}, 2014, pp. 943--954. [Online]. Available: \url{https://doi.org/10.1145/2568225.2568290}
\BIBentrySTDinterwordspacing

\bibitem{ghanbarihadi_omission_2018}
\BIBentryALTinterwordspacing
GhanbariHadi, VartiainenTero, and SiponenMikko, ``\BIBforeignlanguage{EN}{Omission of {Quality} {Software} {Development} {Practices}},'' \emph{\BIBforeignlanguage{EN}{ACM Computing Surveys (CSUR)}}, Feb. 2018, publisher: ACMPUB27New York, NY, USA. [Online]. Available: \url{https://dl.acm.org/doi/10.1145/3177746}
\BIBentrySTDinterwordspacing

\bibitem{paternoster_software_2014}
\BIBentryALTinterwordspacing
N.~Paternoster, C.~Giardino, M.~Unterkalmsteiner, T.~Gorschek, and P.~Abrahamsson, ``Software development in startup companies: {A} systematic mapping study,'' \emph{Information and Software Technology}, vol.~56, no.~10, pp. 1200--1218, Oct. 2014. [Online]. Available: \url{https://www.sciencedirect.com/science/article/pii/S0950584914000950}
\BIBentrySTDinterwordspacing

\bibitem{giardino_software_2016}
\BIBentryALTinterwordspacing
C.~Giardino, N.~Paternoster, M.~Unterkalmsteiner, T.~Gorschek, and P.~Abrahamsson, ``Software {Development} in {Startup} {Companies}: {The} {Greenfield} {Startup} {Model},'' \emph{IEEE Transactions on Software Engineering}, vol.~42, no.~6, pp. 585--604, Jun. 2016, conference Name: IEEE Transactions on Software Engineering. [Online]. Available: \url{https://ieeexplore.ieee.org/abstract/document/7360225}
\BIBentrySTDinterwordspacing

\bibitem{curtis_field_1988}
\BIBentryALTinterwordspacing
B.~Curtis, H.~Krasner, and N.~Iscoe, ``\BIBforeignlanguage{en}{A field study of the software design process for large systems},'' \emph{\BIBforeignlanguage{en}{Commun. ACM}}, vol.~31, no.~11, pp. 1268--1287, Nov. 1988. [Online]. Available: \url{https://dl.acm.org/doi/10.1145/50087.50089}
\BIBentrySTDinterwordspacing

\bibitem{Autowarefoundation}
\BIBentryALTinterwordspacing
Autowarefoundation, ``Autowarefoundation/autoware.universe.'' [Online]. Available: \url{https://github.com/autowarefoundation/autoware.universe}
\BIBentrySTDinterwordspacing

\bibitem{claude}
A.~Askell, Y.~Bai, A.~Chen, D.~Drain, D.~Ganguli, T.~Henighan, A.~Jones, N.~Joseph, B.~Mann, N.~DasSarma, N.~Elhage, Z.~Hatfield-Dodds, D.~Hernandez, J.~Kernion, K.~Ndousse, C.~Olsson, D.~Amodei, T.~Brown, J.~Clark, S.~McCandlish, C.~Olah, and J.~Kaplan, ``A general language assistant as a laboratory for alignment,'' 2021.

\bibitem{lewis_retrieval-augmented_2021}
\BIBentryALTinterwordspacing
P.~Lewis, E.~Perez, A.~Piktus, F.~Petroni, V.~Karpukhin, N.~Goyal, H.~Küttler, M.~Lewis, W.-t. Yih, T.~Rocktäschel, S.~Riedel, and D.~Kiela, ``\BIBforeignlanguage{en}{Retrieval-{Augmented} {Generation} for {Knowledge}-{Intensive} {NLP} {Tasks}},'' Apr. 2021, arXiv:2005.11401 [cs]. [Online]. Available: \url{http://arxiv.org/abs/2005.11401}
\BIBentrySTDinterwordspacing

\bibitem{wei_chain--thought_2023}
\BIBentryALTinterwordspacing
J.~Wei, X.~Wang, D.~Schuurmans, M.~Bosma, B.~Ichter, F.~Xia, E.~Chi, Q.~Le, and D.~Zhou, ``Chain-of-{Thought} {Prompting} {Elicits} {Reasoning} in {Large} {Language} {Models},'' Jan. 2023, arXiv:2201.11903 [cs]. [Online]. Available: \url{http://arxiv.org/abs/2201.11903}
\BIBentrySTDinterwordspacing

\bibitem{yao_react_2023}
\BIBentryALTinterwordspacing
S.~Yao, J.~Zhao, D.~Yu, N.~Du, I.~Shafran, K.~Narasimhan, and Y.~Cao, ``{ReAct}: {Synergizing} {Reasoning} and {Acting} in {Language} {Models},'' Mar. 2023, arXiv:2210.03629 [cs]. [Online]. Available: \url{http://arxiv.org/abs/2210.03629}
\BIBentrySTDinterwordspacing

\bibitem{ezzini_ai-based_2023}
\BIBentryALTinterwordspacing
S.~Ezzini, S.~Abualhaija, C.~Arora, and M.~Sabetzadeh, ``{AI}-based {Question} {Answering} {Assistance} for {Analyzing} {Natural}-language {Requirements},'' in \emph{2023 {IEEE}/{ACM} 45th {International} {Conference} on {Software} {Engineering} ({ICSE})}, May 2023, pp. 1277--1289, iSSN: 1558-1225. [Online]. Available: \url{https://ieeexplore.ieee.org/document/10172663}
\BIBentrySTDinterwordspacing

\bibitem{rodriguez_prompts_2023}
\BIBentryALTinterwordspacing
A.~D. Rodriguez, K.~R. Dearstyne, and J.~Cleland-Huang, ``\BIBforeignlanguage{en}{Prompts {Matter}: {Insights} and {Strategies} for {Prompt} {Engineering} in {Automated} {Software} {Traceability}},'' Aug. 2023. [Online]. Available: \url{https://arxiv.org/abs/2308.00229v1}
\BIBentrySTDinterwordspacing

\bibitem{bert_pl}
\BIBentryALTinterwordspacing
J.~Lin, Y.~Liu, Q.~Zeng, M.~Jiang, and J.~Cleland-Huang, ``\BIBforeignlanguage{en}{Traceability transformed: Generating more accurate links with pre-trained bert models},'' \emph{\BIBforeignlanguage{en}{arXiv:2102.04411 [cs]}}, Feb 2021, arXiv: 2102.04411. [Online]. Available: \url{http://arxiv.org/abs/2102.04411}
\BIBentrySTDinterwordspacing

\bibitem{bert_nl}
\BIBentryALTinterwordspacing
J.~Lin, A.~Poudel, W.~Yu, Q.~Zeng, M.~Jiang, and J.~Cleland-Huang, ``\BIBforeignlanguage{en}{Enhancing automated software traceability by transfer learning from open-world data},'' no. arXiv:2207.01084, Jul 2022, arXiv:2207.01084 [cs]. [Online]. Available: \url{http://arxiv.org/abs/2207.01084}
\BIBentrySTDinterwordspacing

\bibitem{reimers_sentence-bert_2019}
\BIBentryALTinterwordspacing
N.~Reimers and I.~Gurevych, ``Sentence-{BERT}: {Sentence} {Embeddings} using {Siamese} {BERT}-{Networks},'' Aug. 2019, arXiv:1908.10084 [cs]. [Online]. Available: \url{http://arxiv.org/abs/1908.10084}
\BIBentrySTDinterwordspacing

\bibitem{githubGitHubAutowarefoundationautowareuniverse}
``{G}it{H}ub - autowarefoundation/autoware.universe --- github.com,'' \url{https://github.com/autowarefoundation/autoware.universe}, [Accessed 27-06-2024].

\bibitem{dearstyne_supporting_2024}
\BIBentryALTinterwordspacing
K.~R. Dearstyne, A.~D. Rodriguez, and J.~Cleland-Huang, ``Supporting {Software} {Maintenance} with {Dynamically} {Generated} {Document} {Hierarchies},'' Aug. 2024, arXiv:2408.05829 [cs]. [Online]. Available: \url{http://arxiv.org/abs/2408.05829}
\BIBentrySTDinterwordspacing

\bibitem{rodriguez_2024_12574870}
\BIBentryALTinterwordspacing
A.~Rodriguez, ``Requirement set with defects,'' Jun. 2024. [Online]. Available: \url{https://doi.org/10.5281/zenodo.12574870}
\BIBentrySTDinterwordspacing

\end{thebibliography}

\end{document}